\documentclass[pss]{wiley2sp} 

\usepackage{bm}             
\usepackage{url}
\usepackage[utf8]{inputenc}
\usepackage[english]{babel}
\usepackage{graphicx}
\usepackage{subfig}
\usepackage{color}
\usepackage{chemfig}
\usepackage{epstopdf}
\AppendGraphicsExtensions{.tif}

\newcommand{\change}[1]{\textcolor{black}{#1}}

\begin{document}

\title{Materials Informatics for Dark Matter Detection}

  
\author{%
  R. Matthias Geilhufe\textsuperscript{\Ast,\textsf{\bfseries 1}},
  Bart Olsthoorn\textsuperscript{\textsf{\bfseries 1}}
  Alfredo Ferella\textsuperscript{\textsf{\bfseries 4}},
  Timo Koski\textsuperscript{\textsf{\bfseries 6}},
  Felix Kahlhoefer\textsuperscript{\textsf{\bfseries 5}},
  Jan Conrad\textsuperscript{\Ast,\textsf{\bfseries 4}},
  Alexander V. Balatsky\textsuperscript{\Ast, \textsf{\bfseries 1,2,3}},
}

\newcommand{\notes}[1]{\textcolor{red}{#1}}

\mail{e-mail
  \textsf{geilhufe@kth.se, balatsky@hotmail.com, conrad@fysik.su.se}}

\institute{%
  \textsuperscript{1}\,Nordita, KTH Royal Institute of Technology and Stockholm University, Roslagstullsbacken 23, SE-106 91 Stockholm, Sweden\\
  \textsuperscript{2}\,University of Connecticut, 2152 Hillside Road, U-3046 Storrs, CT 06269-3046, USA\\
  \textsuperscript{3}\,Institute for Materials Science, Los Alamos National Laboratory, Los Alamos, NM 87545, USA\\
\textsuperscript{4}\, Oskar Klein Centre for Cosmoparticle Physics, Fysikum Stockholm University, Roslagstullsbacken 21, SE-10961 Stockholm Sweden\\
  \textsuperscript{5}\, Institute for Theoretical Particle Physics and Cosmology (TTK), RWTH Aachen University, D-52056 Aachen, Germany\\
  \textsuperscript{6}\, Department for Mathematics, KTH Royal Institute of Technology, Lindstedtsvägen 25, SE-10044 Stockholm Sweden}


\abstract{
Dark Matter particles are commonly assumed to be weakly interacting massive particles (WIMPs) with a mass in the GeV to TeV range. However, recent interest has shifted towards lighter WIMPs, which are more difficult to probe experimentally. A detection of sub-GeV WIMPs would require the use of small gap materials in sensors. Using recent estimates of the WIMP mass, we identify the relevant target space towards small gap materials (100-10 meV). Dirac Materials, a class of small- or zero-gap materials, emerge as natural candidates for sensors for Dark Matter detection. We propose the use of informatics tools to rapidly assay materials band structures to search for small gap semiconductors and semimetals, rather than focusing on a few preselected compounds. As a specific example of the proposed strategy, we use the organic materials database (omdb.diracmaterials.org) to identify organic candidates for sensors: the narrow band gap semiconductors BNQ-TTF and DEBTTT with gaps of 40 and 38 meV, and the Dirac-line semimetal (BEDT-TTF)$\cdot$Br which exhibits a tiny gap of $\approx$ 50 meV when spin-orbit coupling is included. We outline a novel and powerful approach to search for dark matter detection sensor materials by means of a rapid assay of materials using informatics tools.
}
\maketitle  

\section{Introduction}
Recently, the evidence for the existence of a significant component of matter in the Universe which is not visible by conventional telescopes has become indisputable, e.g. \cite{Bergstrom:2012fi}. To date, the nature of this matter, coined dark matter (DM), is still unknown. Currently, the most popular theoretical model of dark matter introduces a new type of elementary particle created in the early Universe, dubbed Weakly Interacting Massive particle (WIMP). The predicted mass of WIMPs ranges from ~GeV to ~TeV, with an interaction strengths roughly on the weak scale. In recent years, searches for this GeV mass DM have reached sensitivities that probe the current paradigm, e.g. \cite{Arcadi:2017kky,Conrad:2017pms,Buchmueller:2017qhf,Kahlhoefer:2017dnp} , and the absence of positive signals motivate the widening of the search window - in particular to masses below GeV. Generically, there are two classes of particles in this "light dark matter" range: sub-GeV DM from a hidden dark sector with a new force interacting with the standard model or ultra-light DM with mass range from 10$^{-22}$ eV to keV. The arguably most popular example of the latter class is the axion, invoked to solve the apparent absence of CP violation in Quantum Chromo Dynamics \cite{Wilczek:1977pj,Weinberg:1977ma}. See  \cite{Battaglieri:2017aum} for a comprehensive review of models and searches.

The experimental  resolution of the DM puzzle requires detection of the particle in the  wide range of DM mass. The DM search space, especially on the low end of mass, does require one to use a new approach to sensors for DM detection. Detection of the light-mass  DM would necessitate the   use of materials that provide signatures at low scattering energy with high fidelity yet would be impervious to the ambient noise in electronics. 

A promising path to detect DM is given by direct detection, i.e., detecting the recoil of DM particles in a target material by measurement of the energy deposited, as light, charge or heat. This approach is strongly connected to the highly non-trivial task of identifying appropriate materials having the necessary target properties, such as, optimal band gap, chemical stability, large single-crystal sizes, or, specific magnetic and dielectric properties. 

The challenge of DM detection requires interdisciplinary approach where sensor materials are selected from a broad list of candidates based on their utility. Motivated by the exponential growth of computational power and the resulting data, we witness the rapid adoption of functional materials prediction within the framework of materials informatics \cite{rodgers2006materials,ferris2007materials,ortiz2009data}. Here, methods adapted from computer science based on data-mining and machine learning are applied to identify materials with requested target properties. This approach has been successful, e.g., in the prediction of thermoelectric materials \cite{PhysRevX.1.021012}, energy storage materials \cite{mendoza2011design}, Dirac materials \cite{Geilhufe20164,Geilhufe2017}, topological insulators \cite{klintenberg2014computational}, and superconductors \cite{geilhufe2018towards,klintenberg2013possible}.

In this paper we define the target space for DM detection sensor materials and discuss the application of data mining techniques on the example of the organic materials database (OMDB) \cite{borysov2017organic}. The OMDB is an electronic structure database for previously synthesized 3-dimensional organic crystals, freely accessible via \url{https://omdb.diracmaterials.org}. 
First, we discuss the detection of DM particles and determine certain functional requirements of potential sensor materials. Afterwards, we illustrate the mining for sensor materials in the realm of organic materials. We start by searching for nonmagnetic tiny gap materials which are rare in the class of the organics. Afterwards, we outline the potential use of organic Dirac materials, where the relatively small spin-orbit coupling lifts the spin-degeneracy and opens a tiny gap.  We thus provide here, for first time,  the  interdisciplinary challenge and propose a specific solution based on the identification of the target space and large data search for the candidate sensor materials and proposed specific examples of materials for DM detection.  

\begin{figure}[b]
\includegraphics[width=9.5cm]{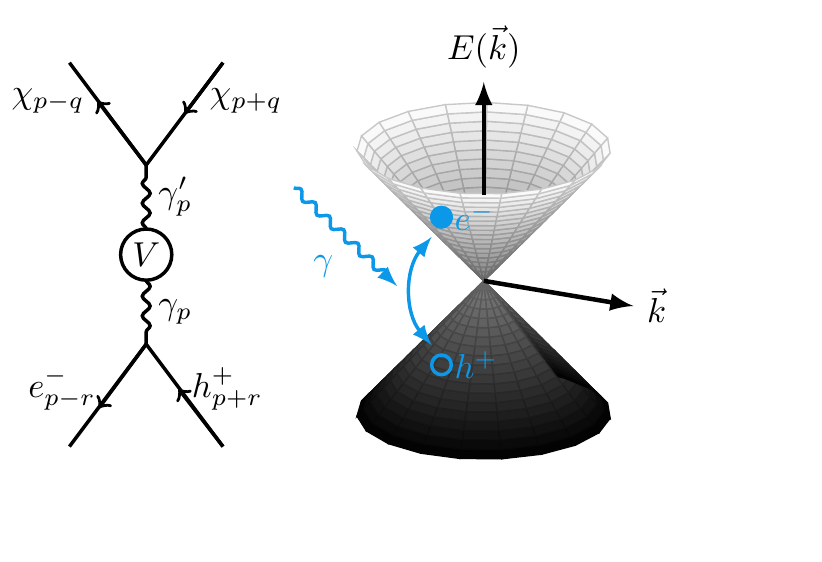}
\caption{Schematic picture of a potential DM particle detection using Dirac materials. An annihilation of a DM particle $\chi$ with its antiparticle creates a dark photon $\gamma'$. The dark photon can be transferred into a light photon $\gamma$ via an undefined process $V$. Finally, the photon can be measured by exciting electron-hole pairs within a Dirac material. \label{Dirac_measuring}}
\end{figure}

\section{The optimal target space for DM detection}
As mentioned above, the basic idea of direct detection of DM is that the interaction of DM with the detector material leads to a detectable energy deposit. The interaction will depend on the type of DM. Elastic scattering of DM particles from electrons or nucleons, DM absorption, couplings of the DM spin are all possible. On the other hand, low mass bosonic DM particles (e.g. axions) would  act collectively as a coherently oscillating classical wave, giving rise to phenomena as time-oscillating electric dipole moments, spin precession detected via nuclear magnetic resonance techniques (dependent on the magnetic properties of target material), e.g. \cite{Budker:2013hfa}. Consequently, the target space for the material will crucially depend on our assumptions on the nature of DM.

Below we give the summary of "constraints" one need to keep in mind when talking about novel detector materials. 

1. {\em Small Gap}. For definiteness, we focus on elastic scattering of DM on electrons. The available recoil energy is limited to $E_\mathrm{max} \approx m_{DM} \, v_\mathrm{max}^2 / 2$ where $v_\mathrm{max} \approx 600 \, \mathrm{km/s} \approx 2 \times 10^{-3} c$ is the escape velocity for DM  bound to the halo, with $c$ being the light velocity. For DM with masses of $\sim$ MeV we find typical maximal recoil energies of $\sim$ eV. Consequently, in  order to achieve sensitivity to light DM (1 MeV/c\textsuperscript{2} and lighter) the deposited energy of the DM particle has to exceed the binding energy of the electrons. Therefore an important prerequisite for the ideal material candidate is to have the lowest possible energy gap (order of meV). On the other hand, the existence of a non-zero energy band gap insures that at sub-kelvin temperatures thermal excitation of electrons to the conduction band is reduced to the minimum, therefore keeping thermal noise sources to the minimum.

2. {\em Shallow slope}. As pointed out in~\cite{DMforDM2}, an essential property of detector materials for the detection of low-mass DM is that the slope of the velocity dispersion relation, i.e.\ the Fermi velocity $v_\mathrm{F}$, is sufficiently small close to the gap. Indeed, for scattering to be kinematically allowed at all, it is necessary that $v_\mathrm{F} \leq v_\mathrm{max}$. Additionally, for optimal sensitivity, $v_\mathrm{F}$ should be in the order of $10^{-4}\dots 10^{-3}$ of the light velocity. In the case of an anisotropic velocity dispersion, at least one direction should have a sufficiently shallow slope.

3. {\em Directional Sensitivity}. Another aspect of low yield DM detection is the possibility to measure the direction of the incoming DM particle. Any detection scheme that would have a good resultion of the direction of incoming DM particles would bring tremendous advantages to our ability to reject backgrounds and establish the detection of DM unambiguously \cite{Mayet:2016zxu}. Directional sensitivity can imply a simple binary response of the detector (e.g. if the material is only responsive to certain direction)  to the possibility to reconstruct the direction of the recoiling particle in three dimensions. 

4. {\em Practicality and Costs}. Apart from material properties that are connected to the physics goal in question there are economical and technical  properties that the material has to fulfill.   The material should be chemically stable (potentially inert, like a noble element). It should be economically and technically possible to synthesize the material in quantities necessary for the experiment.  In addition the material should be free of anomalies or impurities that can alter the desired material parameters or influence the electronic band structure and band gap.  Impurities can also  introduce a source of background.  Currently used detector materials such as Germanium cost about 80,000 Euro per kilogram, which provides a benchmark against which to test the proposed material. 

These criteria represent some, but not all the key elements that constrain the search space for materials. For example detectors focused on axions and larger mass DM particles will have qualitatively different requirements. Therefore, the presented set of criteria can be expanded and modified, depending on the needs. 

\subsection{Read-out}
In order to enhance the extremely small signal from a low-energy electronic recoil induced by light DM scattering or absorption we can likely exploit the Neganov-Luke effect \cite{Neganov:1985khw,Luke:1990ir}. In order to achieve this, a voltage across the target crystal will be applied and the phonons produced in such a process will be detected using a transition edge sensor (TES) or an equivalent sensor like neutron transmutation doped (NTD) germanium detector or a microwave kinetic inductance detector (MKID). It has already been demonstrated by SuperCDMS \cite{Agnese:2018col} that a sensitivity to single electron-hole pair (in a small silicon crystal) can be achieved with such low temperature technologies. Since the thermal gain is inversely proportional to the mean energy required to generate electron-hole pairs (related to the band gap) and considering the very small gap of the materials considered we think that the Naganov-Luke effect could even operate more efficiently than in the conventional semi-conductors, where this effect has been exploited.

Technologies used in quantum computing and quantum sensing  probably can significantly expand possible schemes for single electron-hole sensitivity.

\section{Low band gap materials}
Identifying tiny band gap semiconductors within the realm of organic materials seems to be a tedious task. It has been reported that the expected band gap distribution of randomly chosen organic crystals follows a Wigner-Dyson distribution with a mean band gap in the order of 3 eV \cite{borysov2017organic}. 
Here we will discuss the search results for generic small gap materials and the utility of small gap Dirac materials as a particular class of materials, conducive to small gap formation. 

We performed a search for narrow band gap organic materials using the OMDB, which currently stores electronic band structures and density of states calculated using density functional theory for 26,739 previously synthesized 3-dimensional organic crystals. Computational details for the OMDB can be found in Ref. 
\begin{figure}[b]
\subfloat[\label{bandgap_dist}]{\includegraphics[width=4.2cm]{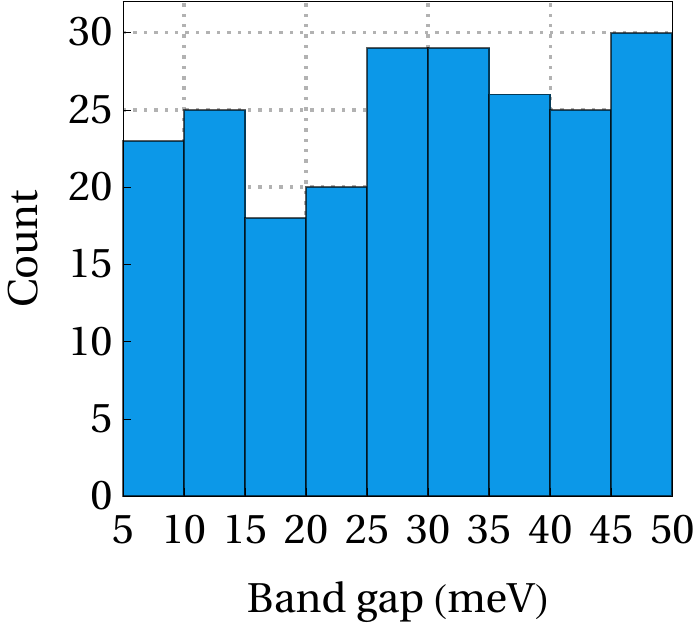}}
\hspace{0.1cm}
\subfloat[]{
\raisebox{0.3cm}{\includegraphics[width=1.5cm]{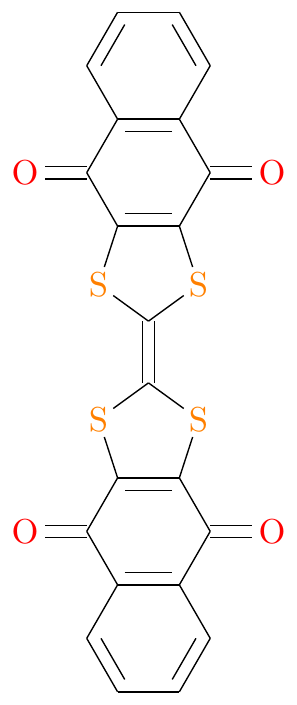}}}\hspace{0.1cm}
\subfloat[]{\raisebox{0.3cm}{\includegraphics[width=2.cm]{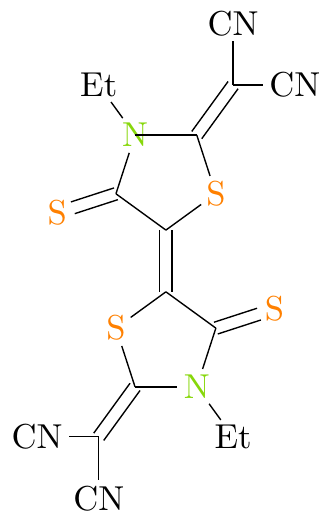}}}
\caption{Data mining for tiny band gap materials within the OMDB. (a) Band gap distribution of tiny gap semiconductors stored within the OMDB. (b) and (c),
molecular structures of tiny band gap materials found within the OMDB. b) BNQ-TTF \cite{C1CE05559C} c) structural modification of DEBTTT \cite{C6CE00772D}. The corresponding band gaps are 40 meV and 38 meV, respectively \cite{omdb21807,omdb24641}.\label{bandgap_materials}}
\end{figure}\cite{borysov2017organic}. 

In our search we found that out of the 26,739 materials stored within the OMDB, only a couple of hundred materials fall into the tiny band gap regime. More precisely, the OMDB stores 225 materials with a band gap in the range of 5 meV and 50 meV (see Fig. \ref{bandgap_dist}) where most of the materials tend to contain magnetic atoms. For magnetic materials the tiny gap is a result of strongly localized electrons exhibiting flat bands and the resulting spin-splitting of these bands due to the magnetic moments. The reliability of these gaps strongly depends on the underlying magnetic structure which can not be determined with high accuracy at the current stage. However, out of the 225 materials, we have identified two nonmagnetic crystals. The first is a triclinic crystal formed of BNQ-TTF (bis(naphthoquinone)-tetrathiafulvalene) molecules \cite{C1CE05559C} exhibiting a PBE band gap of 40 meV. The second material crystallizes in a monoclinic crystal structure formed of structural modifications of the charge transfer salt DEBTTT  given by (E)-3,3'-diethyl-5,5'-bithiazolidinylidene-4,4'-dithione-2,2'-di(dicyanovinylidene) \cite{C6CE00772D}, having a PBE gap of 38 meV. The band structures of the materials can be found within the OMDB \cite{borysov2017organic}, where the respective OMDB-IDs are given by 21807 \cite{omdb21807} and 24641 \cite{omdb24641}. The molecular structures are shown in Figure \ref{bandgap_materials}.

\section{Organic Dirac materials}

Recently Dirac materials have been suggested to be a promising class of materials for DM sensors \cite{DMforDM1,DMforDM2}. Due to their potential for providing very small band-gaps and the directional dependence of DM particles detection, Dirac Materials offer unique advantages.  

A schematic measurement process for DM particle detection is illustrated in Figure \ref{Dirac_measuring}\change{, in the context of an absorption process. Such a process could be realized, when the ordinary photon as well as the dark photon fields are coupled by a kinetic mixing parameter \cite{DMforDM2}}. Dirac materials refer to a class of materials where the quasiparticle excitations can be effectively described by means of a Dirac equation \cite{wehling2014dirac}. Prominent examples comprise of Graphene, topological insulators, Weyl semimetals and Dirac semimetals. These materials are characterized by a linear crossing of energy bands right at the Fermi level, which can be gapped out by symmetry breaking perturbations or spin-orbit coupling (SOC). The latter mechanism is of particular interest in the context of DM detection, as it gives rise to tiny gaps in the meV regime, especially if light elements are involved.

Organic Dirac materials are particularly promising as the typical constituents comprise of light elements such as carbon, hydrogen, oxygen, nitrogen and sulfur. Few well-studied examples for Dirac semimetals exist for inorganic materials. However, the realm of organic Dirac materials remains rather unexplored to date. 
One of the most prominent examples of organic Dirac materials are the quasi 2-dimensional charge transfer salts, where, e.g., $\alpha$-(BEDT-TTF)$_2$I$_3$ was reported to undergo a semiconductor-semimetal phase transition under application of high pressure \cite{katayama2006pressure} or chemical strain \cite{geilhufe_BEDT}. 

\begin{figure}[t]
\subfloat[]{
\hspace{1.3cm}
\includegraphics[width=4.cm]{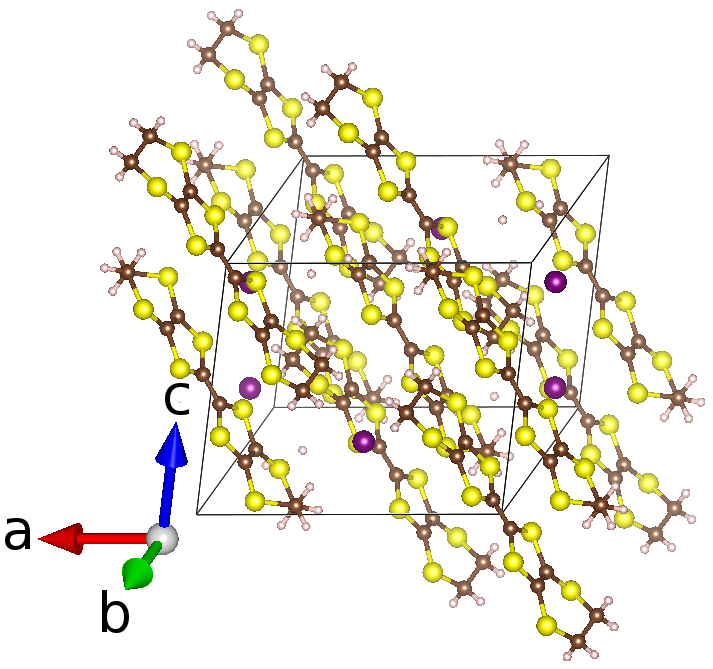}\hspace{0.4cm}
\raisebox{0.3cm}{\includegraphics[width=1.cm]{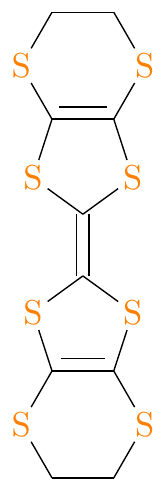}}
}
\\
\subfloat[]{\includegraphics[width=8cm]{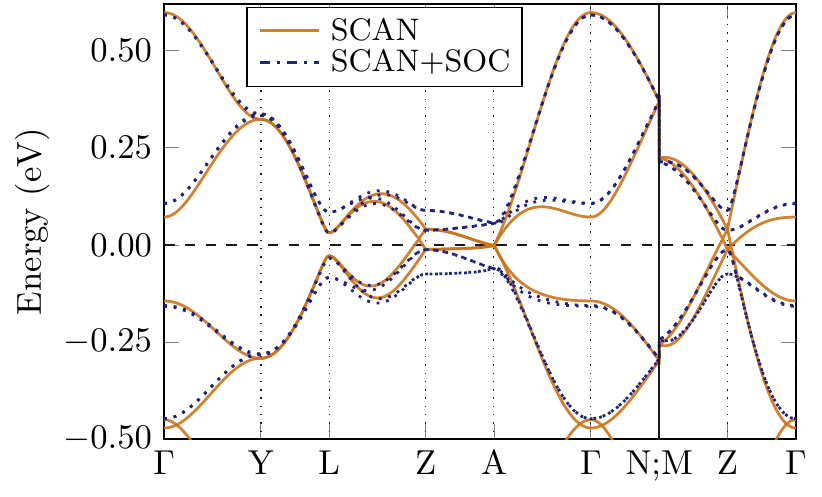}}
\caption{Crystal and electronic structure of the organic Dirac material (BEDT-TTF)$\cdot$Br. Calculations were performed in the framework of the DFT using the SCAN meta-GGA functional. The points in the Brillouin zone denote (in relative units) $\Gamma=(0,0,0)$, $Y=(1/2,1/2,0)$, $L=(1/2,1/2,1/2)$, $Z=(0,0,1/2)$, $A=(0,1/2,1/2)$, $N=(1/2,0,0)$, and $M=(1/2,0,1/2)$. \label{bedt_bands}}
\end{figure}
Data mining  the OMDB we could identify the 3-dimensional organic crystal (BEDT-TTF)$\cdot$Br crystallizing in the monoclinic space group C2/c (SG15) \cite{omdb35335}. The nonsymmorphic glide-mirror symmetry of the material protects nodal lines within the electronic band structure \cite{fang2015topological,yang}. To improve the accuracy of our prediction we used a two-step approach. First we identified the no gap organic material with line nodes. Subsequently we turned on the small spin orbit coupling term to generate a small gap.  We performed ab initio calculations in the framework of the density functional theory (DFT) using a pseudopotential projector augmented-wave method~\cite{hamann1979norm,blochl1994projector,pseudo1,pseudo2}, as implemented in the Vienna Ab initio Simulation Package (VASP)~\cite{vasp2,kresse1999ultrasoft}. We approximated the exchange correlation functional by applying the semilocal meta-GGA functional (SCAN) \cite{sun2015,sun2016accurate} and incorporating Van der Waals corrections according to Tkatchenko and Scheffler \cite{tkatchenko2009accurate}. For the $\vec{k}$-space integration, we chose a $6\times6\times6$ $\Gamma$-centered mesh~\cite{monkhorst1976special} and the precision flag was set to ``normal''. A structural optimization was performed by allowing the ionic positions to change ($\textsc{isif}=2$).  

The calculated band structure of (BEDT-TTF)$\cdot$Br is shown in Figure \ref{bedt_bands}. Without spin-orbit coupling (SOC) the electron filling leads to nodal lines located right at the Fermi level, giving rise to a nodal line semimetallic state. In relative coordinates, the crossings are found in the $k_z=1/2$ plane where spin degenerate bands intersect. At the $A=(0,1/2,1/2)$ point in the Brillouin zone, a crossing of four spin degenerate bands can be seen. To discuss the applicability of the found organic Dirac material as a DM sensor, we have used the calculated band structure to fit the Dirac velocity at various points in the Brillouin zone. Here, we obtained $v_D\approx 4.3\times10^{-4}\,c$ for the crossing along the path $\overline{\mathrm{LZ}}$, $v_D\approx 6.1\times10^{-5}\,c$ for the crossing at the $\mathrm{A}$ point dispersing towards $\mathrm{Z}$, and $v_D\approx 5.0\times10^{-4}\,c$ for the crossing at $\mathrm{A}$ dispersing towards $\mathrm{\Gamma}$. Incorporating SOC, these degeneracies are lifted and a tiny gap in the order of $\approx$ 50 meV can be verified. 

This calculation highlights the combined approach we plan to use in the future to search for relevant functional materials for sensors. 

\section{Conclusions}
Fundamental physics, especially the search for particle DM is starting to enter a phase where available technology rather than theoretical arguments become the driver of experiments. Hence we foresee a growing need for predictive materials capability that would rapidly identify the relevant subset of materials with desired functionalities. In the context of DM detection the narrow band gap materials are of particular interest. We laid out the interdisciplinary approach that is comprised of linked steps: i) we  identify the target space of relevant materials based on specific needs; ii) we used informatics tools  to search the large organic materials database OMDB, given the target space. 

We have followed two directions for mining  to identify narrow band gap materials. First, we identified two non-magnetic molecular crystals with fully occupied molecular orbitals which happen to have a tiny gap in the order of 40 meV. Second, we identified a 3-dimensional organic Dirac-line semimetal (BEDT-TTF)$\cdot$Br. Without SOC, some of the arising crossings at the Fermi level exhibit a Dirac velocity in the order of $4\dots 5\times10^{-4}\,c$, which is the correct order of magnitude for dark matter detection. However, the weak spin-orbit coupling present in the material lifts the spin-degeneracy and opens a gap of approximately 50 meV. \change{According to Ref. \cite{DMforDM2}, an absorption process of dark matter particles can be discussed on the example of kinetically mixed dark photons. The sensitivity for such an absorption process depends on the materials polarization tensor according to the optical theorem, as well as the dark photon mass, acquired by a dark Higgs or Stueckelberg mechanism \cite{AN2013,An20132}. However, a precise calculation of the optical properties of the proposed materials is beyond the scope of this paper.}

Our approach illustrates the potential of the material informatics approach for finding suitable materials for DM detection. Next steps in this effort will include predictions for other materials, e.g inorganic Dirac materials with small gaps,  DM detection sensitivity and an expansion of the target space as well as an investigation of a reliable read out scheme.

\section{Acknowledgement}\label{acknowledgements}
We are grateful to K. Zarembo for useful discussions. This work is supported in part by the Institute for Materials Science at Los Alamos. We are grateful for support from the Swedish Research Council Grant No.~638-2013-9243, the Knut and Alice Wallenberg Foundation, and the Villum Fonden via the Centre of Excellence for Dirac Materials (Grant No.  11744). 
The authors acknowledge computational resources from the Swedish National Infrastructure for Computing (SNIC) at the High Performance Computing Center North (HPC2N).

\bibliographystyle{pss}
\bibliography{mybib.bib}

\end{document}